\documentclass{elex}

\usepackage{graphicx,color}
\usepackage{pgfplots}
\usepackage{algorithm,algorithmic}
\usepackage{subfig}
\usepackage{boxhandler}

\vol{*}
\no{*}

\title{An FPGA-based centralized visible light beacon network\\
}

\author{Duc-Phuc Nguyen$^{\rm a)}$, Dinh-Dung Le$^{}$}

\affiliate{%
	ETIS Laboratory, ENSEA, Paris-Seine University\\
	6 Avenue du ponceau, 95014 Cergy Cedex, FRANCE\\
Graduate School of Science and Technology, Nara Institute of Science and Technology\\
630-0192 Takayama-cho 8916-5, Ikoma-shi, Nara-ken, Japan}

\email{{\rm a)} nguyen.duc-phuc@ensea.fr}

\begin{document}

\maketitle 

\begin{abstract}
Indoor localization systems based on Visible Light Communication (VLC) have shown promising advantages compared with systems based on other wireless technologies. In these systems, many VLC light-emitting diode (LED) anchors are employed in an indoor space in which location identification messages are sent to user devices in small packets. In normal beacon network models, micro-controller (MCU) or low-end system-on-chip (SoC) are often the coordinators which configure messages for one or many VLC-LED bulbs. In this paper, we discuss about processing overload and implementation cost of the two typical models of VLC beacon network in scenarios of a hundred of VLC-LED anchors are taken into account. Finally, an FPGA-based centralized VLC transmitter and its aided Nios II-based system has been introduced to enhance the performance of the VLC beacon network. Besides, due to the centralized processing, our system model is considered to be more cost-efficient than the dedicated-processor-based models.   
\end{abstract}



\section{Background}
\label{intro}
Indoor Location-Based Services (ILBS) are getting more attractions from researchers and industry due to their realistic applications. Generally, there are many wireless technologies applied in current ILBS such as WiFi, Bluetooth, Ultrasound, radio frequency identification (RFID), Zigbee and so on \cite{wang2018light,luo2017indoor}. Recently, visible light communication (VLC)-based indoor positioning systems (IPS) which possess promising characteristics of high bandwidth, energy-efficient, long lifetime and cost-efficient, are becoming strong candidates in ILBS market \cite{luo2017indoor,chowdhury2018comparative,khan2017visible,zhuang2018survey}. Consequently, VLC-based IPSs are now appeared in indoor public spaces, factories, logistics, shopping and healthcare facilities \cite{zhuang2018survey}. Recently, Japan Electronics and Information Technology Industries Association (JEITA) has standardized the visible light beacon system in which unique identification (ID) messages are transmitted from each VLC-LED bulb for purposes such as identifying objects and locations \cite{jeita2013}. At the user's device, photodiode-based or smartphone camera-based receivers decode the received light signals to retrieve the transmitted information \cite{zhuang2018survey,le2018log}. Finally, localization algorithms (e.g., proximity and triangulation) are executed by the firmware of user portable devices to estimate the location of the receiver. 

\begin{figure}[H]
	\centering
	
	\subfloat[VLC beacon network based on dedicated control boards (dedicated-processor-based).]{
		\label{Fig_1a}
		\includegraphics[width=0.7\textwidth]{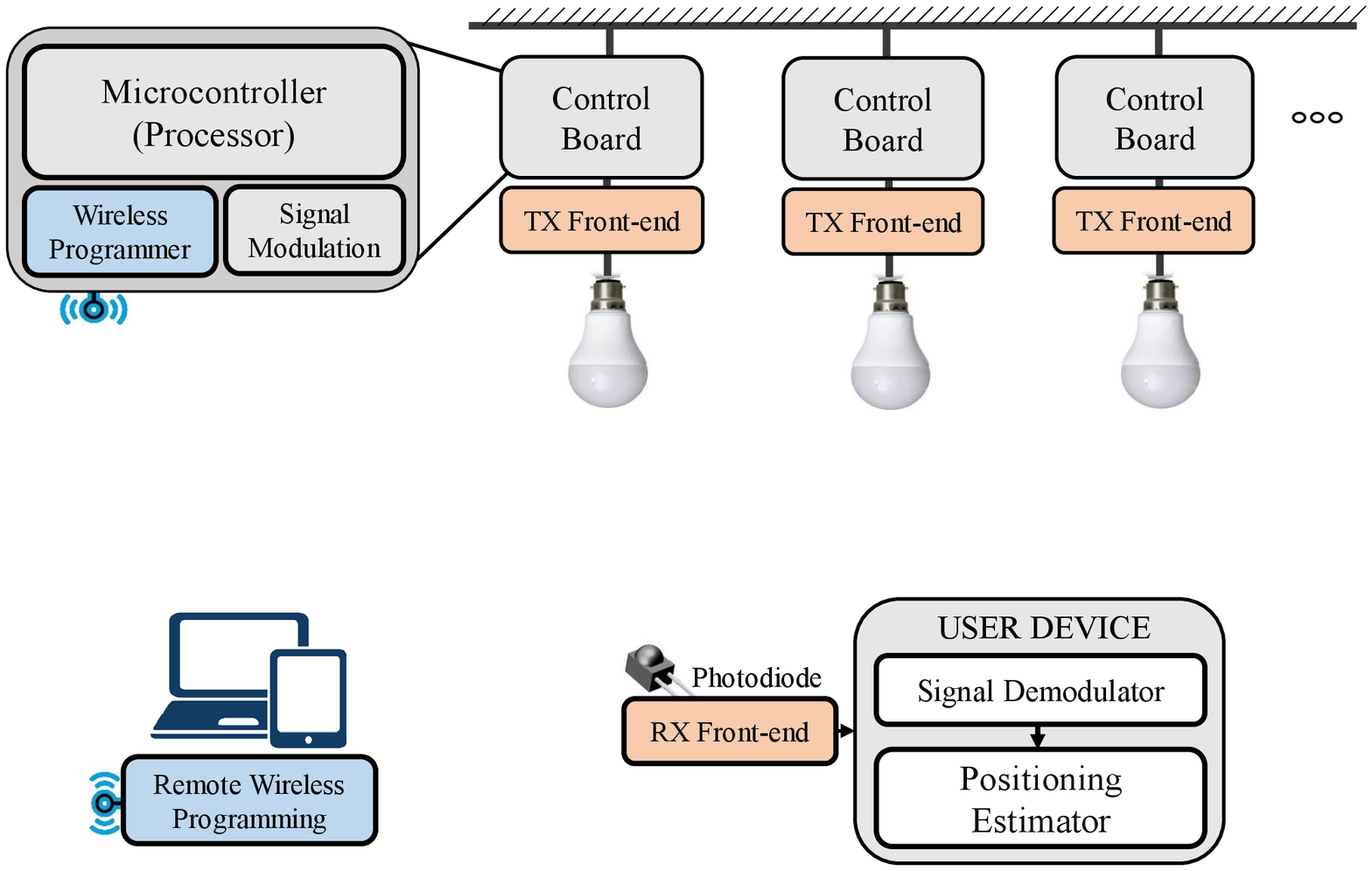} } 
	
	\subfloat[VLC beacon network based on a central control board (central-processor-based)]{
		\label{Fig_1b}
		\includegraphics[width=0.7\textwidth]{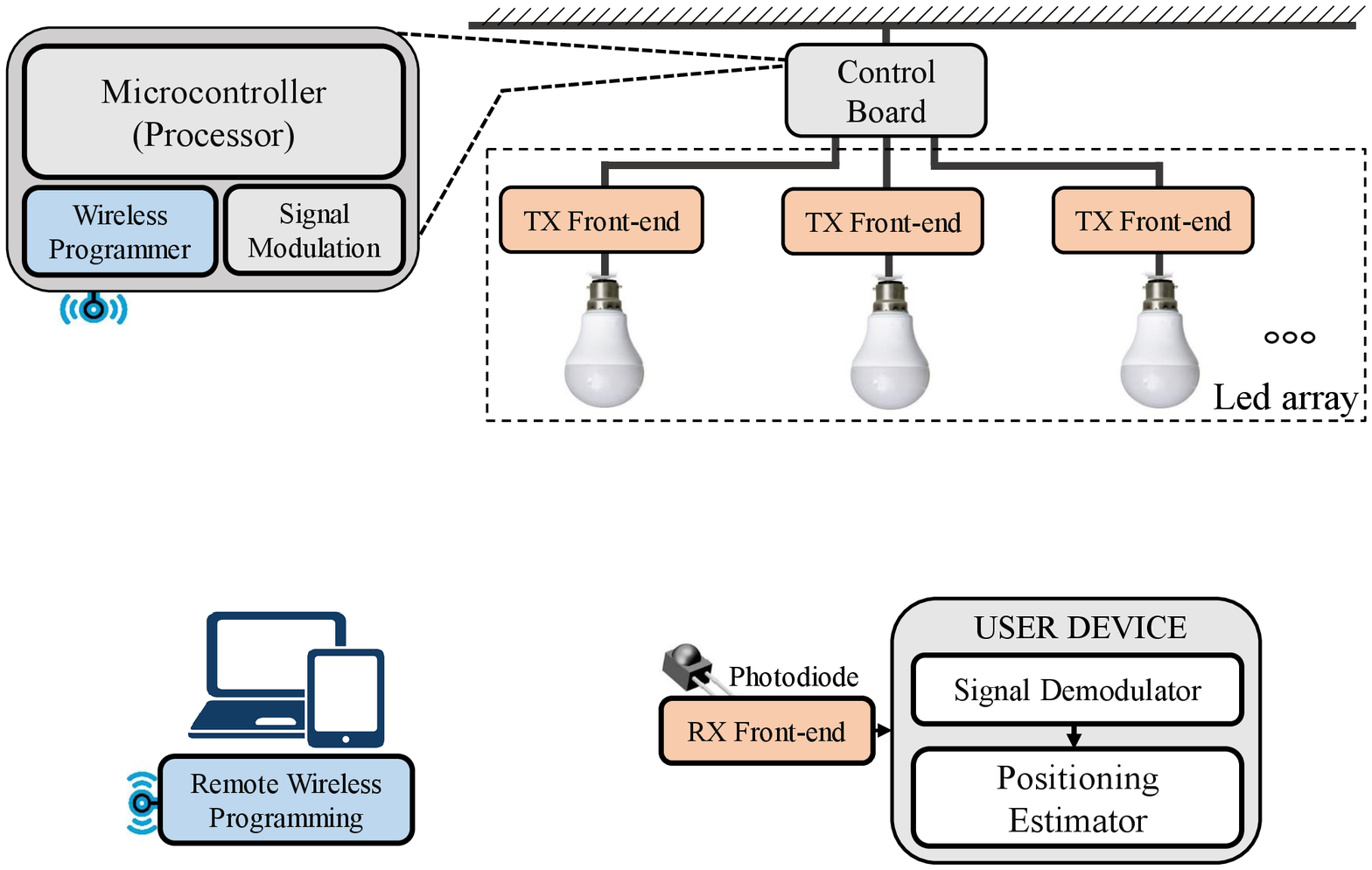} } 	
	\caption{Two typical models of VLC-based IPS}
	\label{Fig_1}
	
\end{figure}

Fig.\ref{Fig_1a} and Fig.\ref{Fig_1b} shows two typical models of VLC-based IPSs. In beacon network presented at Fig.\ref{Fig_1a}, each VLC-LED anchor is controlled by one control unit which we called dedicated-processor-based model \cite{wu2017smartvlc,qiu2016let,li2014epsilon}. Whereas, Fig.\ref{Fig_1b} shows a beacon network in which LED array is controlled by one central processor, we called this model as central-processor-based model \cite{zhuang2018survey,hassan2015indoor}. In these two beacon network models, VLC transmitter's procedures which including forward error correction (FEC) and run-length limited (RLL) encoding \cite{khan2017visible}, are mainly processed by a firmware program on a low-end embedded processor. Besides, due to multiple LEDs are installed in an indoor space, the \textit{Signal Modulation} block is required to execute multiplexing protocols, such as frequency division multiplexing (FDM) and time division multiplexing (TDM); this block ensures that signals from different LEDs can be differentiated at the receiver \cite{hassan2015indoor}. Furthermore, an optional part of the VLC transmit (TX) package is the wireless programmer which helps configure the firmware on the micro-controller or low-end SoC remotely \cite{tuan2018demonstration,wu2017smartvlc,qiu2016let}. At the receiver, signal demultiplexing and positioning algorithms are processed by a firmware program on user's portable device.

\section{Problem and related works}
\label{problem}

\begin{figure}[htb]
	\begin{center}
		\includegraphics[width=10cm]{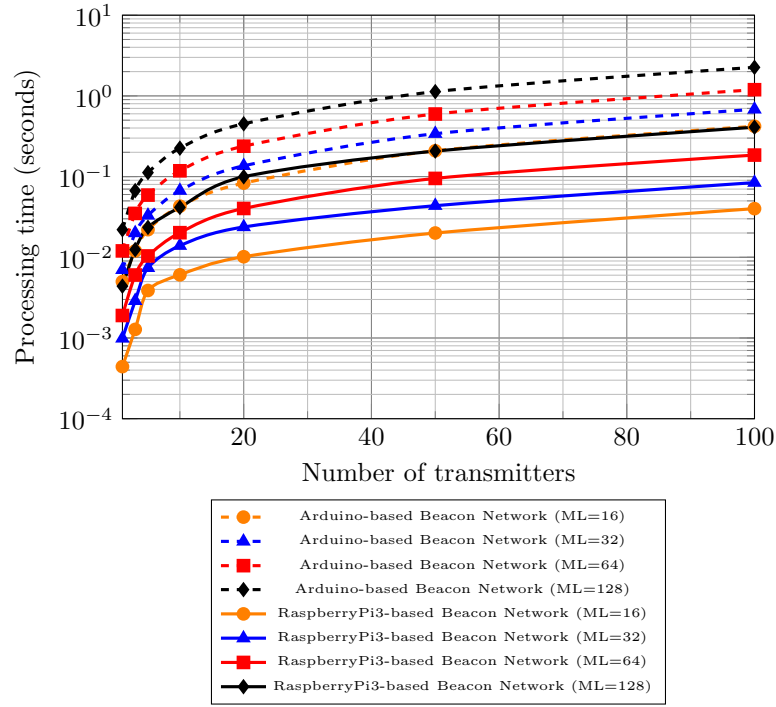}
	\end{center}
	\caption{Evaluation of processing delay of central-processor-based beacon networks. Arduino Uno and Raspberry Pi 3 boards are selected for evaluations}
	\label{Fig_2}
\end{figure}

According to \cite{zhuang2018survey}, system implementation cost is the first priority when considering the commercial availability of a VLC-based IPS design. In the dedicated-processor-based model (Fig.\ref{Fig_1a}), considering this model is applied in a large building with hundreds or thousands of roof VLC-LED bulbs. In this scenario, the implementation cost increases linearly because each control board is dedicated for only one VLC-LED anchor \cite{tuan2018demonstration,wu2017smartvlc,qiu2016let,li2014epsilon}. Moreover, each VLC-LED anchor takes more space to integrate the control board and TX front-end into the same VLC-LED package. Due to these reasons, the dedicated-processor-based model should be further considered to be applied in real commercial systems.  

On the other hand, considering the central-processor-based model is applied with hundreds of LED anchors, and long wires are required to route from the central control board to LED bulbs via VLC TX front-ends (Fig.\ref{Fig_1b}). In this scenario, transmitter's essential procedures include encoding of FEC and RLL codes \cite{tran2016per,khan2017visible, nguyen2018vlsi} are processed sequentially at the central control board's firmware. Next, encoded messages are modulated and be forwarded to VLC TX front-ends. Although, FEC and RLL encoding are often not heavy computations. However, computation efforts increase when more LED anchors are installed in the beacon network. Hence, the central-processor-based model potentially causes critical delays in processing time of FEC, RLL or modulation routines. Moreover, due to the limited storage capability of most low-end embedded processor, the network scalability could be restricted.

\begin{algorithm}
	\caption{The algorithm used in evaluation of central-processor-based beacon network. The transmitter's procedures include Polar encoding and Mancherter RLL encoding; or Polar encoding and 4B6B RLL encoding}
	\label{alg:the_alg1}
	\begin{algorithmic}[1]
		\renewcommand{\algorithmicrequire}{\textbf{Input:}}
		\renewcommand{\algorithmicensure}{\textbf{Output:}}
		\REQUIRE message array $mes[0:K-1]$, Frozen bit location index array $d[ ]$
		\ENSURE  $outMan[0:2N-1]$ (Manchester encode); or $out4b6b[0:2N/3-1]$ (4B6B encode)
		\\ \textit{Initialisation} : \\
		+ $k$ : the number of transmitters in beacon network.\\ 
		+ $N$ : codeword length, $N = 2^n$ \\
		+ $bitIndex$ = 0; $x$ = 0 
		\FOR {$num = 0$ to $k-1$}
		\FOR {$c = 0$ to $N-1$}
		\IF {$d[c]$ is a frozen bit}
		\STATE $polarEn[c] \leftarrow  0$
		\ELSE
		\STATE $polarEn[c] \leftarrow  mes[bitIndex]$
		\STATE $bitIndex \leftarrow bitIndex + 1 $
		\ENDIF
		\ENDFOR
		
		\FOR {$i = 0$ to $n-1$}
		\STATE $b \leftarrow 2^{n-i}$
		\STATE $nb \leftarrow 2^{i}$	
		\FOR {$j = 0$ to $nb-1$}
		\STATE $base \leftarrow j*b$
		\STATE $bdiv2 \leftarrow b/2$
		\FOR {$t = 0$ to $bdiv2-1$}
		\STATE $polarEn[base + t] \leftarrow modulo_2(polarEn[base + t] + polarEn[base + t + bdiv2])$
		
		\ENDFOR
		\ENDFOR
		\ENDFOR	
		
		\FOR {$z = 0$ to $2N-1$} 
		\IF {$polarEn[z/2] = 1$}
		\STATE $outMan[z] \leftarrow 1 ; outMan[z+1] \leftarrow 0$
		\ELSE
		\STATE $outMan[z] \leftarrow 0 ; outMan[z+1] \leftarrow 1$
		\ENDIF
		\STATE $z = z + 2$
		
		\ENDFOR
		\FOR {$z = 0$ to $2N/3-1$} 
		\STATE $out4b6b[z+5,z+4,z+3,z+2,z+1,z] = 4B6B Lookup table ($ \\${polarEn[x+3],polarEn[x+2],polarEn[x+1],polarEn[x]})$
		\STATE $z = z + 6$ ;  $x = x + 4$ 
		
		\ENDFOR
		
		\ENDFOR
		\RETURN $outMan[0:2N-1]$; or $out4b6b[0:2N/3-1]$
	\end{algorithmic} 
\end{algorithm}

To confirm the statement, we have evaluated the consumed memories and processing delays of sequentially executing VLC transmitters' procedures on Arduino Uno and Raspberry Pi 3 boards which are such two representatives between many low-end, low-cost MCUs and SoCs available on the market. Indeed, Reed-Solomon (RS) and convolutional codes (CC) are defined as FEC solutions in three operating modes of VLC transmitters \cite{rajagopal2012ieee}. Besides, Manchester and 4B6B codes are also defined as main candidates for RLL codes in low-speed PHY I operating mode \cite{rajagopal2012ieee, nguyen2018hardware}. Recently, Polar-code-based FEC solutions and soft-decoding theories of RLL codes have been introduced to increase the performance of VLC systems \cite{wang2018decoding,le2018joint,wang2017dimming,fang2017efficient, le2018log}. In this paper, two Polar-code-based transmitters are selected to evaluate in two beacon network models. In the first transmitter, Polar encoding concatenated with Manchester RLL encoding are implemented. Whereas, 4B6B is selected as the RLL encoding solution for the second transmitter. Also, On-off keying (OOK) is selected as the modulation scheme because of its simplicity; and dimming support functions are not covered in the evaluation. 

\begin{table}[ht]
	\begin{center}
		\caption{Amount and percentages of global variables consumed to execute one transmitter's routines on Arduino Uno} \label{globalvariables}
		\begin{small}
			\begin{tabular}{ccc}
				\hline
				&     \begin{tabular}{@{}c@{}}\textbf{Polar + Manchester} \\ (Code rate = 1/4)\end{tabular}   &   \begin{tabular}{@{}c@{}}\textbf{Polar + 4B6B} \\ (Code rate = 1/3)\end{tabular}   \\
				\hline
				ML=16, CL=32 		 &	452 bytes (22\%)	&  	422 bytes (20\%)   \\
				\hline 
				ML=32, CL=64 		 &	644 bytes (31\%)    & 	582 bytes (28\%) \\
				\hline 
				ML=64, CL=128	     &  1028 bytes (50\%)   &   902 bytes (44\%)   \\
				\hline
				ML=128, CL=256	     & 	1796 bytes (87\%)   &   1542 bytes (75\%)  \\
				\hline
			\end{tabular}
		\end{small}
	\end{center}
\end{table}
The Polar-code-based VLC transmitters' procedures are described in Algorithm \ref{alg:the_alg1}. We have implemented two VLC transmitters on Arduino Uno and Raspberry Pi 3 boards to evaluate the processing delays in different message lengths (ML) (ML = 16-bit, 32-bit, 64-bit, 128-bit), corresponding to different codeword lengths (CL) of Polar code (CL = 32-bit, 64-bit, 128-bit and 256-bit). The evaluation results are shown at Fig.\ref{Fig_2}. We have found that processing delay increases linearly when number of transmitters increases in the beacon network; this is a critical point in any ILBS where users always expect real-time responses. Besides, Table \ref{globalvariables} summarizes the amount of consumed global variables of transmitter's procedures. Due to the limited dynamic memory of 2048 bytes, Arduino Uno consumes 22\%, 31\%, 50\%, 87\% of memory resource to storage all global variables (GV) of Manchester-based transmitter's routines; with codeword lengths vary from 32, 64, 128 to 256 respectively. Also, in case 4B6B RLL encoding is applied with Polar encoding, smaller percentages of dynamic memory consumption are reported. However, in case of ML = 128 and CL = 256, the memory consumption rates of 87\% and 75\% might cause some instabilities when the system is in operation.

Evaluation results of processing time and consumed memory of central-processor-based VLC-LED beacon network (Fig.\ref{Fig_1b}) have shown an critical processing delay when number of transmitters increases in beacon network. Besides, limited storage capabilities of low-end MCUs create barriers for this model to be applied in reality. On the other hand, due to the high-cost of implementation, the dedicated-processor-based VLC beacon network (Fig.\ref{Fig_1a}) is also not a effective solution for real commercial systems. Therefore, in this paper, we introduce an FPGA-based centralized transmitter and its aided on-chip system, which can solve problems of processing delay and memory overload; and this solution is expected to support for a larger VLC beacon network.       
        
\section{Proposed system}
\label{proposed}

\begin{figure}[htb]
	\begin{center}
		\includegraphics[width=10cm]{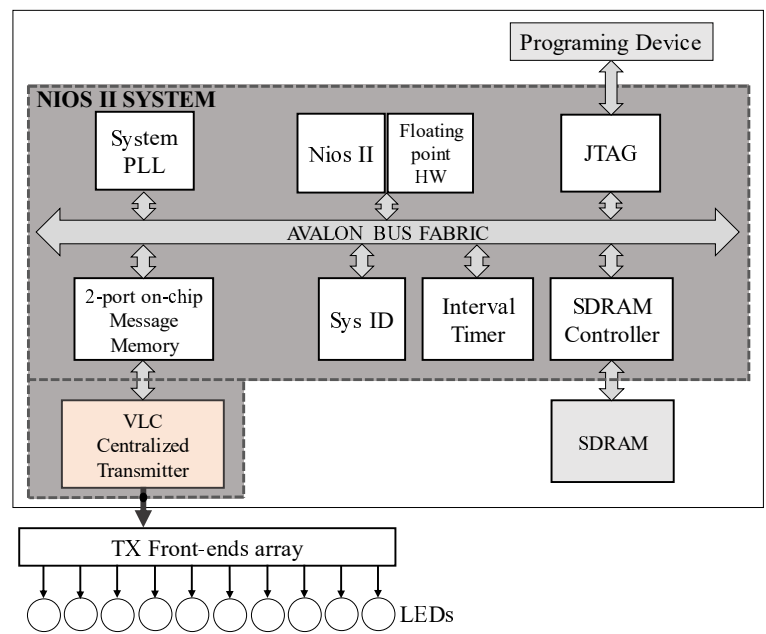}
	\end{center}
	\caption{VLC centralized transmitter and the aided Nios-II-based system}
	\label{Fig_3}
\end{figure}
\begin{figure}[htb]
	\begin{center}
		\includegraphics[width=12cm]{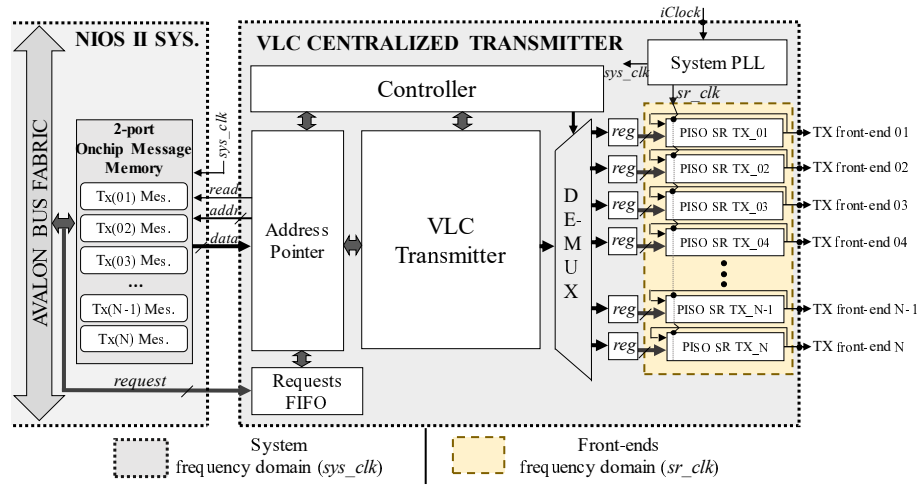}
	\end{center}
	\caption{Hardware architecture of the proposed centralized transmitter.}
	\label{Fig_4}
\end{figure}

To solve problems mentioned in Section \ref{problem}, we have proposed a beacon network based on a centralized VLC transmitter and its aided Nios II system on FPGA. Specifically, FPGA-based centralized beacon network enables all messages could be processed at the central FPGA-based transmitter before encoded messages are passed to TX front-ends. Altera DE2-115 board which features Cyclone IV FPGA chip is selected to implement our system. Indeed, due to the parallel operating capabilities of the FPGA-based logic circuits and many pins are available on common FPGA devices. Many VLC front-ends can be covered by our FPGA-based central processing node.

\subsection{A Nios II system for re-configuration}
\label{niossystem}
An overview of reconfiguration system is presented briefly in Fig.\ref{Fig_3}. The Nios II system includes some basic blocks of any typical system on programmable chip (SoPC). Specifically, the system includes one Nios II soft-processor with specialized hardware for floating point calculations; a JTAG block which connect to programing, debug and monitoring device. Besides, an 64MB SDRAM off-chip is used to store the firmware, an interval timer helps measure the processing time of program. Besides, a 2-port on-chip memory is proposed to storage all uncoded messages of all front-ends in beacon network. In our system, 128-bit is the maximum size of each message, while 100 is the number of front-ends selected for evaluation. Therefore, the 2-port message memory with 1600 bytes can be extended to serve a larger beacon network because of plentiful availability of FPGA's on-chip memory bits. In addition, the system is configured to operate at frequency of 50 Mhz (\textit{sys\_clk}) which is created from internal phase-locked loop (\textit{System PLL}). 

\subsection{VLC Centralized Transmitter}
\label{architecture}
Fig.\ref{Fig_4} describes the hardware architecture that we have implemented for the centralized transmitter. An explanation of this architecture can be divided into 5 parts.
\subsubsection{Clock domains}
\label{clock}
There are two clock domains in the design: system clock (\textit{sys\_clk} = 50 Mhz), and clock for shift registers (\textit{sr\_clk} = 100 Khz). These two clocks are created from the \textit{System PLL} with the reference clock (\textit{iClock} = 50 Mhz). In addition, the VLC TX front-ends transmit information at the same frequency with shift registers' frequency (\textit{sr\_clk}), and this frequency could be adjusted following requirements of the expected VLC system.   

\subsubsection{Requests FIFO and Address Pointer}
\label{command_resolve}
In general cases, the VLC-based IPS initially configures new ID messages for all LED bulbs when IPS is first settled in some indoor spaces. However, there are some scenarios that messages are determined to send to some appointed LED bulbs. In these scenarios, there is no need to update messages for all LED anchors. Therefore, the our centralized transmitter stores all write requests sent to 2-port message memory. Each request is a combination of signals: write request, address, and data to write. In our design, each request is an 136-bit signal which includes 128-bit of message. A (first-in first-out) FIFO buffer is used to storage requests. The \textit{Address Pointer} checks the busy status of the \textit{VLC Transmitter}; then it reads one request stored in FIFO and execute the request. The requests are executed by issuing read request (\textit{read}) and address signal (\textit{addr}) to the second interface of the 2-port on-chip message memory (Fig.\ref{Fig_4}). Next, when \textit{Address Pointer} achieves message from the memory interface (\textit{data}), the acquired message is forwarded to VLC transmitter for FEC and RLL encoding procedures. 

\subsubsection{VLC Transmitter}
\label{vlc_transmitter}
In Section \ref{problem}, we have introduced two VLC transmitters that we have implemented for evaluation. Particularly, the first transmitter includes procedures of Polar and Manchester RLL encoding; while the second transmitter procedures are the concatenation of Polar encoding with 4B6B RLL encoding. These two transmitters are recently mentioned in \cite{wang2018decoding,le2018joint,wang2017dimming,fang2017efficient}. In these two transmitter, the Polar encoders are implemented with architecture inherited from our previous work \cite{nguyen2018hardware, nguyen2018vlsi}. As mentioned earlier, dimming control function is not implemented in these two transmitters. The reason is, although puncturing and compensation symbols (CSs) are purely simple routines; however, these procedures require many storage bits on variable memory, which has been demonstrated about its limitation in Section \ref{problem}. Hence, dimming control block is neglected in our hardware implementation for a fair comparison with Arduino-based model.   
  
\subsubsection{De-multiplexer and registers}
\label{mux_reg}
After the message is processed by \textit{VLC transmitter} block, it is expected to be distributed to appointed front-end. Therefore, a de-multiplexer (\textit{DE-MUX}) determines the front-end registers that the message should be passed, and the (\textit{DE-MUX}) is controlled by a memory-read address that \textit{Address Pointer} has issued. Also, we have implemented loop parallel-input serial-output shift-registers (\textit{PISO SRs}) which can repeat the encoded messages while there is no new messages come to front-ends. \textit{PISO SRs} are operated in front-end frequency domain (\textit{sr\_clk}). Besides, buffering registers (\textit{reg}) are inserted between \textit{DE-MUX} and \textit{PISO SRs} to buffer the message.        

\subsubsection{Controller}
\label{controller}

The \textit{Controller} block handles the operation of VLC centralized transmitter. Specifically, it controls the start and finish of the \textit{VLC Transmitter}; acquires the memory-read address from the \textit{Address Pointer} and gives control signals to \textit{DE-MUX}.

\section{Experimental results} 
The proposed architecture of VLC centralized transmitter is described by synthesizable \textit{Verilog HDL} language.  \textit{ModelSim} is used as the verification tool. The Nios II system is created with the helps of \textit{Platform Designer} tool. \textit{Nios II Software for Eclipse} is used for firmware programing and debugging. Our system is synthesized by \textit{Intel's Quartus II}. Table \ref{niossynthesis} summarizes the synthesis report of Nios II system on Cyclone IV FPGA device. It can be noticed that the Nios II system only consumes 1\% of memory bits of Cyclone IV FPGA; this means that on-chip message memory can be further extended to serve for a larger number of TX front-ends in a larger beacon network. Besides, Table \ref{synthesis} shows the synthesis report of the Manchester-based and 4B6B-based VLC centralized transmitters. In particular, due to a better code rate, the transmitter based on Polar and 4B6B RLL encoding consumes less logic elements (LE), look-up tables (LUT) and registers than the Manchester-based transmitter does. However, due to the storage of 4B6B mapping tables, amount of consumed memory bits of 4B6B-based transmitter is larger than the Manchester-based one. Besides, our FPGA-based centralized transmitters occupy 102 pins in the total of 529 pins of Cyclone IV FPGA (19\%). Indeed, we have just implemented an architecture which only supports 100 front-ends for evaluation; however, the availability of unused pins enables more front-ends can be supported. Additionally, both transmitters can achieve maximum throughputs higher than 600 Mbps; therefore, our FPGA-based centralized transmitter could be potentially applied in high-speed VLC systems. Furthermore, Table \ref{resource} shows resource summary of the centralized transmitters and their components. It can be seen that the de-multiplexer, buffering registers and PISO shift registers occupy most of logic cells in both centralized transmitters. However, instead of using logic cells to implement shift registers, we can utilize embedded memory bits which are still abundant in Cyclone IV FPGA to reduce the total logic cells of the system. 

\begin{table}[ht]
	\begin{center}
		\caption{FPGA synthesis report of the Nios II system} \label{niossynthesis}
		\begin{small}
			\begin{tabular}{cc}
				\hline
				&    \textbf{Nios II System}    \\
				\hline
				Device 			& 	Cyclone IV FPGA   \\
				\hline
				Model 		 &	1200 mV, $0^{o}C$	 \\
				\hline 
				Fmax 		 &	80.25 Mhz   \\
				\hline 
				LE/LUT		   &  12166/114480 (11\%)	 \\
				\hline
				Registers 	& 	7479 	\\
				\hline
				Memory bits &	54713/3981312 (1\%)	 \\
				\hline
				Embedded Multiplier &	15/532 (3\%)	 \\
				\hline
				Total PLLs &	1/4 (25\%)	 \\
				\hline
			\end{tabular}
		\end{small}
	\end{center}
\end{table}

\begin{table}[ht]
	\begin{center}
		\caption{FPGA synthesis report of the FPGA-based centralized transmitters} \label{synthesis}
		\begin{small}
			\begin{tabular}{ccc}
				\hline
				FEC, RLL 			& 	Polar, Manchester &	 Polar, 4B6B   \\
				\hline
				Device 			& 	Cyclone IV FPGA &	 Cyclone IV FPGA   \\
				\hline
				Number of front-ends 	& 	100 &	 100   \\
				\hline
				Model 		 &	1200 mV, $0^{o}C$	&  1200 mV, $0^{o}C$	 \\
				\hline 
				Fmax 		 &	76.13 Mhz   & 69.69 Mhz \\
				\hline 
				Code length		&	256	&  256 \\
				\hline
				Code rate		    &	1/4	& 1/3  \\
				\hline 
				LE/LUT		   &  91518/114480 (80\%)	 &  78823/114480 (69\%)	 \\
				\hline
				Registers 	& 	91004 & 78274	\\
				\hline
				Memory bits &	277/3981312 ($<$1\%)	& 6425/3981312 ($<$1\%)\\
				\hline
				Total pins &	102/529 (19\%) &	102/529 (19\%) \\
				\hline
				Total PLLs &	1/4 (25\%)	&1/4 (25\%) \\
				\hline
				Latency \\($sys\_clk$ domain)  &	14 clock cycles	& 14 clock cycles\\
				\hline
				Maximum throughput &	694.8 Mbps & 	630.8 Mbps \\
				\hline
			\end{tabular}
		\end{small}
	\end{center}
\end{table}

\begin{table}[ht]
\begin{center}
\caption{Resource summary of component blocks of the FPGA-based centralized transmitters; Manchester-based transmitter (left index) and 4B6B-based transmitter (right index)} \label{resource}
\begin{small}
\renewcommand\footnoterule{}
\begin{tabular}{cccccc}
\hline
\textbf{Instance} &\textbf{Logic Cells}&  \textbf{Registers}  & \textbf{ Mem.$^\ddag$} &  \textbf{LUT/Reg.$^\dag$}		   \\
\hline
Request FIFO 			& 	$55 | 54$	   &	$40 | 40$ 		   &	$224 | 224$			    & 		$31 | 30$   	   \\
\hline
Address Pointer 		&	$11 | 11$ 	   &  	$11 | 11$ 		   &	$0 | 0$			    & 		$11 | 11$					   \\
\hline 
Controller			    &	$17 | 17$	   &	$15 | 15$		   &   	$0 | 0$		        &		$15|15$					   \\
\hline
VLC Transmitter		   &  $1388 | 1418$	   &   $1338 | 1408 $   	   &   $53 | 6201$  			   &  		$908 | 714$				   \\
\hline
De-multiplexer\\Front-end regs\\PISO Shift reg. & 	$90062 |77326$		&   $89600 |76800$	&  	$0|0$	&  $54759|41849$ \\
\hline
Total &		$91518 |78823$		 & $91004 | 78274 $ & $277 | 6425 $   & $55708 | 42615 $ 												   \\
\hline
\end{tabular}
\rule{0in}{1.2em}$^\dag$\scriptsize LUT/Registers Logic Cells\\
\rule{0in}{1.2em}$^\ddag$\scriptsize Embedded memory bits\\
\end{small}
\end{center}
\end{table}

\begin{figure}[htb]
	\begin{center}
		\includegraphics[width=10cm]{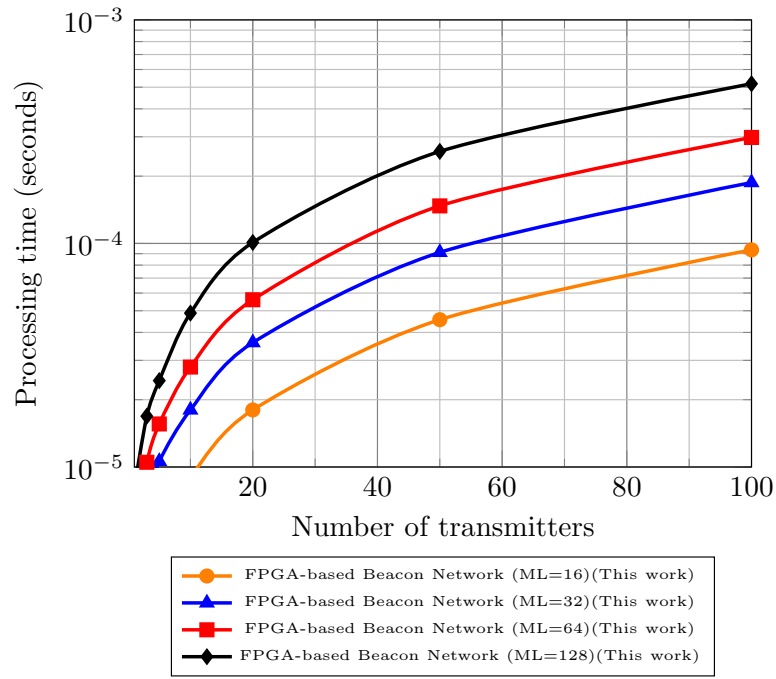}
	\end{center}
	\caption{Processing time of proposed FPGA-based centralized beacon network}
	\label{Fig_5}
\end{figure}

\begin{table}[ht]
	\begin{center}
		\caption{Processing delay enhancement (at ML = 128)} \label{delaygain}
		\begin{small}
			\begin{tabular}{ccc}
				\hline
			No. of Transmitter	&    FPGA/Arduino Gain 	    &    FPGA/Raspberry Gain    \\
				\hline
				1			 	& 		2729				& 	  	548			    	\\
				\hline
				3 		 		&	  	3969				& 		738					\\
				\hline 
				5 				&	 	4609				& 		966 				\\
				\hline 
				10 				&	 	4610				&		850					 \\
				\hline 
				20 				&	  	4465				&		985					 \\
				\hline 
				50				&	    4375				& 		802					\\
				\hline 
				100 	 		&	   	4359				& 		789					\\
				\hline 
			\end{tabular}
		\end{small}
	\end{center}
\end{table}

Fig.\ref{Fig_5} shows the processing delay evaluation of our FPGA-based centralized beacon network. Besides, Table \ref{delaygain} summarizes the processing delay improvement of FPGA-based beacon network. Specifically, improvements of maximum 4610 times and 966 times are reported for processing delay of FPGA-based solution compared evaluation results on Arduino Uno and Raspberry Pi 3, respectively (Fig.\ref{Fig_2} and Fig.\ref{Fig_5}).     
\section{Conclusion}
In this paper, we have introduced an FPGA-based centralized beacon network. Our proposal includes a hardware architecture for the centralized VLC transmitter which can process messages for all TX front-ends in beacon network; and a Nios II-based system to control the messages and operation of the beacon network. Experimental results have shown that our system can improve the processing delay of the central-processor-based beacon networks remarkably. Besides, our FPGA-based model can be extended to serve for a large beacon network which includes many VLC-LED bulbs due to the abundant availability of embedded memory bits and FPGA's pins. Moreover, compared with beacon networks which based on dedicated embedded processors, our FPGA-based centralized system is expected to reduce the implementation cost of the commercial VLC-based positioning systems.

\section*{Acknowledgments}
This work was supported by JSPS KAKENHI Grant Number JP16K18105.

\end{document}